\documentclass{article}
\usepackage{spconf,amsmath,graphicx,hyperref}
\usepackage{booktabs}
\usepackage{xcolor}
\usepackage{multirow}

\title{AudioGenie-Reasoner: A Training-Free Multi-Agent Framework for Coarse-to-Fine Audio Deep Reasoning}
\name{%
  Yan Rong$^{1}$,
  Chenxing Li$^{2}$,
  Dong Yu$^{2}$,
  Li Liu$^{1}$\sthanks{Corresponding Author: avrillliu@hkust-gz.edu.cn}
}
\address{%
  $^{1}$ The Hong Kong University of Science and Technology (Guangzhou)\\
  $^{2}$ Tencent AI Lab
}

\begin{document}

\maketitle

\begin{abstract}
Audio deep reasoning is a challenging task that requires expert-level perception, multi-step logical inference, and the integration of contextual knowledge. However, existing models suffer from a gap between audio perception and reasoning abilities due to the lack of training data with explicit reasoning chains and the absence of mechanisms for active exploration and iterative refinement.
To address these challenges, we propose \textbf{AudioGenie-Reasoner (AGR)}, the \textbf{first} unified training-free multi-agent system that coordinates perception and reasoning over an evolving chain of textual evidence. 
Our key idea is a paradigm shift that transforms audio deep reasoning into complex text understanding task from a new perspective, thereby unlocking the full potential of large language models. 
Specifically, the design of AGR mimics the human coarse-to-fine cognitive process. It first transforms the input audio into a coarse text-based document. Then, we design a novel proactive iterative document refinement loop, featuring tool-augmented routes and specialized agents, to continuously search for missing information and augment the evidence chain in a coarse-to-fine manner until sufficient question-related information is gathered for making final predictions. 
Experimental results show that AGR achieves state-of-the-art (SOTA) performance over existing open-source audio deep reasoning models across various benchmarks. The code will be available at \url{https://github.com/ryysayhi/AudioGenie-Reasoner}.
\end{abstract}

\begin{keywords}
Audio Deep Reasoning, Multi-Agent, Training-Free, Large Language Models, Iterative Refinement
\end{keywords}

\section{Introduction}
Audio deep reasoning~\cite{su2025audio,ma2025mmar,sakshimmau} is a challenge task in audio understanding, requiring expert-level perception, multi-step logical inference, and the integration of contextual knowledge to interpret complex acoustic scenes. This technology has many applications in our daily life, such as embodied intelligence~\cite{liu2024caven} and autonomous systems~\cite{nie2024reason2drive}.

\begin{figure}[t]
    \centering
    \includegraphics[width=1.00\linewidth]{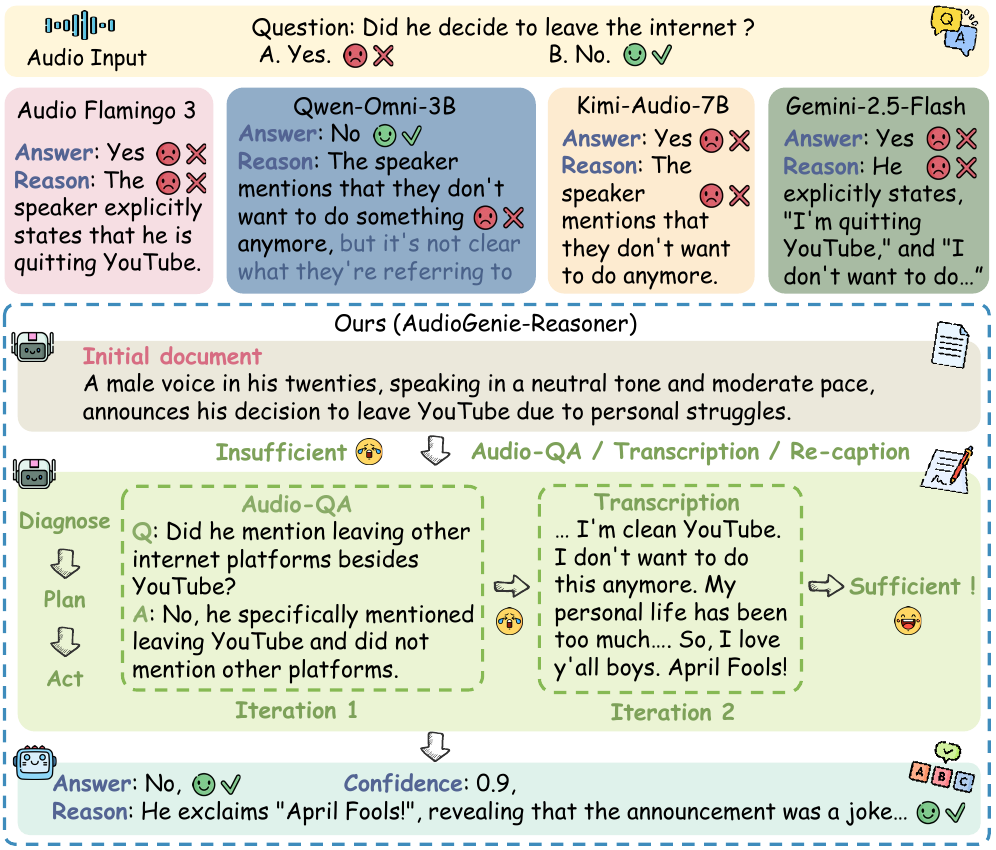}
    \vspace{-1.7em}
    \caption{Performance comparison of AudioGenie-Reasoner with other audio reasoning models. Our framework excels in providing correct answers and valid reasoning.}
    \label{fig:network}
    \vspace{-1.6em}
\end{figure}

\begin{figure*}[t]
    \centering
    \includegraphics[width=0.96\linewidth]{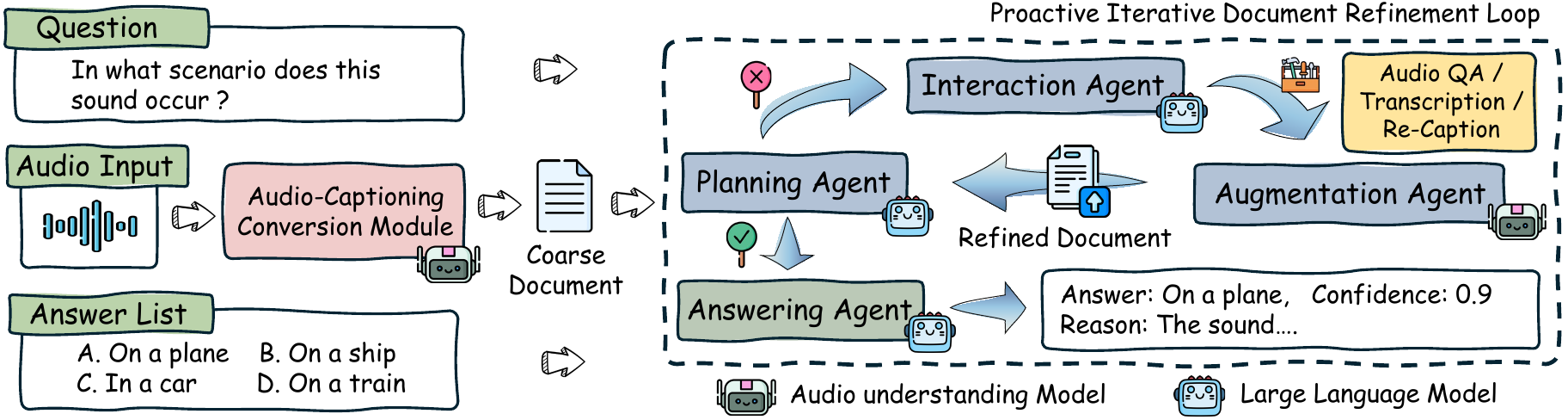}
    \vspace{-0.8em}
    \caption{The multi-agent architecture of AudioGenie-Reasoner. Specialized agents for planning, interaction, and augmentation collaborate within an iterative loop to refine a coarse initial caption into an evolving textual evidence chain.}
    \label{fig:network}
    \vspace{-0.8em}
\end{figure*}

In the literature, great progress in audio reasoning has been achieved by prior works~\cite{xie2025audio,goel2025audio,ding2025kimi,ghosh2024gama}. However, audio deep reasoning remains a significant challenge. The recently proposed audio deep reasoning benchmark, MMAR~\cite{ma2025mmar}, reveals the poor performance of existing audio models. On this challenging benchmark, many open-source models fail to achieve an accuracy better than random guessing, reflecting a gap between the abilities of audio perception and the cognitive reasoning. This gap stems from two fundamental challenges:
\textbf{Firstly}, existing models are hindered by the lack of training data with explicit reasoning chains. Constructing high-quality, step-by-step reasoning annotations for audio is resource-intensive. Lacking this fine-grained supervision, most Audio Large Language Models (ALLMs) are trained on simpler objectives like audio-text alignment~\cite{elizalde2023clap} or direct question-answering~\cite{lipping2022clotho}. In particular, when it comes to complex scenarios with mixed audio sources (\textit{e.g.}, speech, music, and sound effects), their reasoning capabilities degrade sharply.
\textbf{Secondly}, current methods lack a mechanism for active exploration and iterative refinement. Models typically function as passive information receivers, generating answers based on a single pass of perceptual results. This static, single-pass process prevents them from diagnosing evidence gaps, planning to acquire missing information, or progressively deepening their understanding. As a result, they are ill-equipped to handle complex problems that require multi-step and in-depth analysis.

To address above two challenges, we propose \textbf{AudioGenie-Reasoner (AGR)}, a novel unified training-free multi-agent system (MAS) that coordinates perception and reasoning over an evolving chain of textual evidence. Our core design mimics the human coarse-to-fine cognitive process: forming an initial general understanding, conducting a detailed examination of relevant cues based on the specific query, and finally drawing a conclusion from sufficient evidence.
Specifically, \textbf{for the first challenge}, instead of directly training heavy audio-reasoning models, we introduce a paradigm shift that transforms audio deep reasoning into a complex text understanding task. This transformation decouples perception from cognition, elegantly bypassing the need for vast audio-specific reasoning data and unlocking the full potential of Large Language Models (LLMs).
\textbf{For the second challenge}, instead of a conventional single-pass pipeline, we introduce a proactive iterative document refinement loop, driven by tool-augmented routes and specialized agents. This process empowers the model to dynamically find the potential missing information and augment these information within language space. Through this ``diagnose-plan-act” loop, the model is transformed from a passive information receiver into an active, self-improving investigator.

In summary, the main contributions of this work are as follows:
\textbf{(1)} A unified training-free MAS, named AudioGenie-Reasoner, which coordinates perception and reasoning over an evolving chain of textual evidence, is proposed. To the best of our knowledge, this is the first exploration of MAS in audio deep reasoning.
\textbf{(2)} We establish a coarse-to-fine cognitive framework that transforms audio reasoning into a text understanding task, featuring a novel proactive iterative document refinement loop to dynamically search for missing information and augment the evidence chain.
\textbf{(3)} Experimental results show that AGR achieves SOTA performance over existing open-source models across various audio deep reasoning benchmarks.


\begin{table*}[t]
\centering
\caption{Comparison with SOTA methods on MMAU-mini. The `/' separates results from raw outputs \textit{vs.} those after GPT-4o post-processing (see Implementation Details). Best performances are highlighted in bold, while second-best are underlined.}
\label{result_mmau_mini}
\resizebox{\linewidth}{!}{
    \begin{tabular}{c|ccc|ccc|c}
    \toprule
    Methods& Sound & Music & Speech & Easy & Medium & Hard & Avg \\
    \midrule 
    Random Guess & 49.25 / 49.25 & 30.24 / 30.24 & 39.94 / 39.94 & 33.93 / 33.93 & 43.70 / 43.70 & 36.44 / 36.44 & 39.80 / 39.80 \\
    \midrule 
    Gemini-2.5-Flash~\cite{vertexai_gemini_2_5_flash_docs} & \underline{74.77} / \underline{76.58} & \underline{65.27} / 65.57 & 72.97 / 75.58 & \underline{64.29} / \underline{65.62} & 75.93 / 76.66 & 66.10 / 70.19 & 71.00 / 71.90 \\
    Gemini-2.0-Flash~\cite{vertexai_gemini20flash_docs_2025} & 73.27 / 73.27 & 64.97 / 64.97 & \textbf{78.38} / \textbf{78.38} & 62.95 / 62.95 & \textbf{77.22} / \textbf{77.22} & \textbf{69.49} / \textbf{69.49} & \underline{72.20} / \underline{72.20}\\
    Gemini-2.0-Flash-Lite~\cite{vertexai_gemini_2_0_flash_lite_docs_2025} & 69.97 / 69.97 & 65.27 / 65.27 & 74.17 / 74.47 & 60.27 / 60.27 & 74.07 / 74.07 & \underline{69.07} / \underline{69.48} & 69.80 / 69.90 \\
    \midrule
    MiDashengLM-7B~\cite{dinkel2025midashenglm}& 66.37 / 69.67 & 58.98 / 58.98 & 61.56 / 62.16 & 53.12 / 53.12 & 68.89 / 70.37 & 55.93 / 58.05 & 62.30 / 63.60 \\
    Audio Flamingo 3~\cite{goel2025audio}& 74.76 / \textbf{76.88} & 60.18 / 61.08 & 60.96 / 63.06 & 58.04 / 59.82 & 70.19 / 71.30 & 61.02 / 63.98 & 65.30 / 67.00 \\
    Audio Flamingo 3 (T)~\cite{goel2025audio}& 69.97 / 74.47 & 59.28 / \textbf{67.37} & 44.74 / 61.26 & 56.70 / 63.84 & 61.85 / 74.07 & 50.42 / 56.78 & 58.00 / 67.70 \\
    Audio-Reasoner~\cite{xie2025audio} & 32.13 / 66.97 & 41.02 / 63.77 & 34.23 / 57.06 & 43.75 / 61.61 & 30.51 / 64.81 & 34.81 / 58.47 & 35.80 / 62.60 \\
    Qwen2.5-Omni-3B~\cite{xu2025qwen2_5omni} & 73.57 / 73.87 & 60.78 / 60.78 & 63.66 / 64.56 & 57.14 / 57.14 & 70.93 / 71.30 & 63.14 / 63.98 & 66.00 / 66.40 \\
    Audio Flamingo 2-0.5B~\cite{ghosh2025audio} & 26.43 / 47.15 & 17.96 / 35.93 & 15.32 / 27.93 & 24.11 / 36.61 & 16.10 / 38.33 & 19.81 / 34.32 & 19.90 / 37.00 \\
    Audio Flamingo 2-1.5B~\cite{ghosh2025audio} & 42.34 / 50.15 & 35.63 / 46.71 & 34.53 / 37.54 & 36.16 / 41.52 & 34.75 / 48.33 & 39.26 / 39.83 & 37.50 / 44.80 \\
    Audio Flamingo 2-3B~\cite{ghosh2025audio} & 62.46 / 63.96 & 50.60 / 55.09 & 39.34 / 47.15 & 49.11 / 52.23 & 53.52 / 58.70 & 46.19 / 50.85 & 50.80 / 55.40 \\
    Kimi-Audio-7B-Instruct~\cite{ding2025kimi} & 59.46 / 74.17 & 42.51 / 58.38 & 61.56 / 66.07 & 44.64 / 56.70 & 59.26 / 71.48 & 52.97 / 63.14 & 54.50 / 66.20 \\
    \midrule
    \multirow{2}{*}{AudioGenie-Reasoner}& \textbf{75.08} / 75.08  & \textbf{66.17} / \underline{66.17} & \underline{76.58} / \underline{76.58} & \textbf{69.20} / \textbf{69.20} & \underline{76.67} / \underline{76.67} & 66.53 / 66.53 & \textbf{72.60} / \textbf{72.60} \\
    & \textcolor{blue!70}{(+8.7)} / \textcolor{blue!70}{(+5.4)} & \textcolor{blue!70}{(+7.2)} / \textcolor{blue!70}{(+7.2)} & \textcolor{blue!70}{(+15.0)} / \textcolor{blue!70}{(+14.4)} & \textcolor{blue!70}{(+16.1)} / \textcolor{blue!70}{(+16.1)} & \textcolor{blue!70}{(+7.8)} / \textcolor{blue!70}{(+6.3)} & \textcolor{blue!70}{(+10.6)} / \textcolor{blue!70}{(+8.5)} & \textcolor{blue!70}{(+10.3)} / \textcolor{blue!70}{(+9.0)}\\
    \bottomrule
    \end{tabular}
}
\vspace{-0.9em}
\end{table*}

\section{Our Method}
Our framework's design is founded on two core innovations. First, we introduce a paradigm shift that transforms the audio reasoning problem into a text-based understanding task, thereby decoupling perception from cognition. Second, we design a proactive multi-agent loop for iterative evidence refinement, turning the system into an active investigator. The overall architecture is illustrated in Figure~\ref{fig:network}.

\subsection{Paradigm Shift: From Audio Reasoning to Text Understanding}

Instead of attempting to build a data-hungry audio reasoning model, we transform the audio reasoning task into a complex text understanding problem. This is achieved by initially converting the raw input audio $A$ into a coarse-grained textual document $D_{0}$:
\begin{equation}\small
D_{0}=\mathcal{F}_{\text {caption}}(A),
\end{equation}
where $\mathcal{F}_{\text {caption }}(\cdot)$ is an audio-captioning module implemented with a powerful ALLM.

This initial transformation is the foundation of our paradigm shift. It decouples the system's perceptual abilities, which are handled by the ALLM, from its cognitive reasoning, which is governed by LLM-based agents in the subsequent steps. By doing so, we elegantly bypass the need for specialized audio-reasoning datasets and instead unlock the vast, pre-existing reasoning capabilities of LLMs. The resulting document, $D_{0}$, serves as the initial state of an evolving evidence chain, forming the textual foundation upon which all subsequent reasoning and refinement will be performed.

\subsection{Proactive Iterative Document Refinement Loop}
To bridge the gap between the coarse initial description and the fine-grained details required for complex queries, we introduce a proactive iterative refinement loop. This loop is coordinated by a team of specialized agents that collaborate to progressively enrich the initial document into a comprehensive evidence chain. At its core, the loop operates iteratively: it first assesses the current evidence, then plans and executes actions to augment it with missing information via tool-augmented routes. This process repeats until the evidence is deemed sufficient for a confident answer. 

\textbf{Planning Agent.}
Each iteration begins with the planning agent $\mathcal{F}_{\text {plan }}(\cdot)$, which assesses if the current document contains sufficient evidence to confidently answer the question:
\begin{equation}\small
\left(s, H_{i+1}\right)=\mathcal{F}_{\text {plan}}\left(Q, L, D_{i}, H_{i}\right),
\end{equation}
where $Q$ is the question, $L$ is the answer list, $D_{i}$ is the document at iteration $i$, and $H_{i}$ is the analysis history. The agent returns a status flag $s \in\{ \text {Sufficient}, \text {Insufficient} \}$. If the evidence is insufficient, the history is updated to $H_{i+1}$ with an analysis of the information gap.

\textbf{Interaction Agent.}
If the status $s$ is insufficient, the interaction agent $\mathcal{F}_{\text {interact }}(\cdot)$ formulates a plan to acquire the missing information:
\begin{equation}\small
P=\mathcal{F}_{\text {interact}}\left(D_{i}, H_{i+1}\right),
\end{equation}
where $P$ is a structured augmentation plan. The plan outlines one of three tool-based actions: audio question-answering, guided re-captioning, or automatic speech recognition.

\textbf{Augmentation Agent.}
The augmentation agent $\mathcal{F}_{\text {Aug }}(\cdot)$ executes the plan $P$ by invoking the specified tool to generate new evidence $E_{\text {new }}=\mathcal{F}_{\text {Aug}}\left(P\right)$ and integrate it into the document:
\begin{equation}\small
D_{i+1}=D_{i} \oplus E_{\text {new}},
\end{equation}
where the $\oplus$ operator denotes the integration of $E_{\text {new }}$ into the existing document $D_{i}$. The enriched document  $D_{i+1}$ is then passed back to the planning agent for the next iteration.

\textbf{Answering Agent.}
Once the iterative refinement loop concludes (\textit{i.e.}, when  $s = \text{Sufficient}$) or the maximum number of iterations is reached, the answer agent $\mathcal{F}_{\text {answer }}(\cdot)$ generates the final output from the enriched document $D_{f}$:
\begin{equation}\small
\left(A^{*}, S_{c}, R\right)=\mathcal{F}_{\text {answer}}\left(D_{f}, Q, L\right),
\end{equation}
where $A^{*}$, $S_{c}$, and $R$ represent the final selected answer, the associated confidence score, and a detailed textual rationale explaining the reasoning process, respectively.

\begin{table*}[t]
\centering
\caption{Comparison with SOTA methods on MMAR. The `/' separates results from raw outputs \textit{vs.} those after GPT-4o post-processing (see Implementation Details). So, Mu, and Sp denote Sound, Music, and Speech, respectively. Best performances are highlighted in bold, while second-best are underlined.}
\label{result_mmar}
\resizebox{\linewidth}{!}{
    \begin{tabular}{c|ccc|cccc|c}
    \toprule
    Methods& Sound & Music & Speech & So-Mu & So-Sp & Mu-Sp & Sn-Mu-Sp &  Avg \\
    \midrule
    Random Guess & 27.74 / 27.74 & 24.58 / 24.58 & 35.38 / 35.38 & 18.18 / 18.18 & 24.63 / 24.63 & 28.00 / 28.00 & 13.64 / 13.64 & 28.18 / 28.18 \\
    \midrule 
    Gemini-2.5-Flash~\cite{vertexai_gemini_2_5_flash_docs} & \textbf{56.13} / \textbf{57.42} & 39.11 / \underline{48.04} & \textbf{76.92} / \textbf{79.23} & 45.45 / 45.45 & \underline{73.40} / \textbf{75.37} & \textbf{68.00} / \textbf{74.67} & 54.55 / 54.55 & \underline{63.43} / \textbf{67.07} \\
    Gemini-2.0-Flash~\cite{vertexai_gemini20flash_docs_2025} & \underline{52.90} / \underline{52.90} & \textbf{53.07} / \textbf{53.07} & \underline{71.15} / \underline{71.15} & \textbf{100} / \textbf{100} & \textbf{73.89} / \underline{73.89} & \underline{66.67} / \underline{68.00} & \textbf{63.64} / \textbf{63.64} & \textbf{64.86} / \underline{64.97} \\
    Gemini-2.0-Flash-Lite~\cite{vertexai_gemini_2_0_flash_lite_docs_2025} & 52.26 / 52.26 & \underline{45.25} / 45.25 & 66.54 / 66.92 & \underline{72.73} / \underline{72.73} & 66.01 / 66.01 & 66.67 / 66.67 & 54.55 / 54.55 & 59.56 / 59.67 \\
    \midrule
    MiDashengLM-7B~\cite{dinkel2025midashenglm}& 43.23 / 43.87 & 40.22 / 40.22 & 51.15 / 51.15 & 18.18 / 18.18 & 45.81 / 45.81 & 57.33 / 57.33 & 36.36 / 36.36 & 46.19 / 46.30 \\
    Audio Flamingo 3~\cite{goel2025audio}& 45.81 / 47.10 & 31.84 / 32.40 & 53.85 / 54.23 & 27.27 / 27.27 & 46.31 / 47.29 & 54.67 / 56.00 & 45.45 / 45.45 & 45.97 / 46.74 \\
    Audio Flamingo 3 (T)~\cite{goel2025audio}& 41.94 / 52.26 & 25.70 / 32.40 & 39.23 / 49.62 & 18.18 / 27.27 & 44.83 / 54.19 & 42.67 / 52.00 & 18.18 / 31.82 & 37.79 / 47.18 \\
    Audio-Reasoner~\cite{xie2025audio} & 26.45 / 39.35 & 20.67 / 35.75 & 24.23 / 39.62 & 36.36 / 54.55 & 28.57 / 44.33 & 38.67 / 48.00 & 27.27 / 31.82  & 26.30 / 40.55 \\
    Qwen2.5-Omni-3B~\cite{xu2025qwen2_5omni} & 50.97 / 50.97 & 46.37 / 46.37 & 48.85 / 48.85 & 27.27 / 27.27 & 51.72 / 51.72 & 61.33 / 61.33 & 45.45 / 45.45 & 50.06 / 50.06 \\
    Audio Flamingo 2-0.5B~\cite{ghosh2025audio} & 11.61 / 21.94 & 6.70 / 16.20 & 13.85 / 22.69 & 9.09 / 9.09 & 15.27 / 26.11 & 17.33 / 22.67 & 18.18 / 22.73 & 12.71 / 21.88 \\
    Audio Flamingo 2-1.5B~\cite{ghosh2025audio} & 21.29 / 25.81 & 20.11 / 29.05 & 24.62 / 28.85 & 9.09 / 9.09 & 18.23 / 21.67 & 26.67 / 30.67 & 27.27 / 31.82 & 21.77 / 26.74 \\
    Audio Flamingo 2-3B~\cite{ghosh2025audio} & 39.35 / 43.23 & 27.37 / 31.84 & 36.15 / 38.85 & 36.36 / 36.36 & 28.57 / 30.05 & 29.33 / 30.67 & 31.82 / 36.36 & 32.60 / 35.47\\
    Kimi-Audio-7B-Instruct~\cite{ding2025kimi} & 49.03 / 50.32 & 32.96 / 37.99 & 52.69 / 56.15 & 18.18 / 36.36 & 56.65 / 61.58 & 52.00 / 60.00 & 36.36 / 45.45 &  48.18 / 52.60 \\
    \midrule
    \multirow{2}{*}{AudioGenie-Reasoner}& 49.68 / 49.68 & 43.26 / 43.26 & 69.23 / 69.23 & 45.45 / 45.45 & 64.53 / 64.53 & 65.33 / 65.33 & \underline{59.09} / \underline{59.09} & 58.85 / 58.85 \\
    & \textcolor{blue!70}{(+6.5)} / \textcolor{blue!70}{(+5.8)} & \textcolor{blue!70}{(+3.0)} / \textcolor{blue!70}{(+3.0)} & \textcolor{blue!70}{(+18.1)} / \textcolor{blue!70}{(+18.1)} & \textcolor{blue!70}{(+27.3)} / \textcolor{blue!70}{(+27.3)} & \textcolor{blue!70}{(+18.7)} / \textcolor{blue!70}{(+18.7)} & \textcolor{blue!70}{(+8.0)} / \textcolor{blue!70}{(+8.0)} & \textcolor{blue!70}{(+22.7)} / \textcolor{blue!70}{(+22.7)} & \textcolor{blue!70}{(+12.7)} / \textcolor{blue!70}{(+12.6)} \\
    \bottomrule
    \end{tabular}
}
\vspace{-1.0em}
\end{table*}

\section{Experimental Results}
\subsection{Experimental Setup}
\textbf{Datasets.} We evaluate our framework on two well-known audio deep reasoning benchmarks: MMAU-mini~\cite{sakshimmau} and MMAR~\cite{ma2025mmar}. MMAU-mini consists of 1,000 closed-form questions covering three audio types: sound, music, and speech. MMAR is a more challenging benchmark that includes not only single audio types but also various mixtures of them. Since the audio data for MMAR is not directly provided, we successfully collected 905 samples after filtering for inaccessible data due to issues like expired links.

\textbf{Implementation Details.} We select MiDashengLM-7B~\cite{dinkel2025midashenglm} and GPT-4o-2024-08-06~\cite{openai_gpt4o_system_card_2024} as the ALLM and LLM in our framework, respectively. Whisper-Turbo~\cite{radford2023robust} is employed as the transcription model in our tool-based actions. The max number of iterations is set to three. 
For the evaluation metric, we follow the methodology of MMAU and MMAR, comparing the model's prediction with the ground truth using regular expressions and string matching. To handle cases where some ALLMs produce semantically correct but improperly formatted answers, \textbf{we use GPT-4o-2024-08-06 to post-process the raw outputs}. This step normalizes the generated text by mapping it to the corresponding answer in the predefined list (\textit{e.g.}, mapping a free-form response like ``The final answer is C" to the third item in the choice list), ensuring a fair and accurate evaluation.

\begin{table}[t]
\centering
\vspace{-0.8em}
\caption{Results of ablation studies on different model components. Best performances are highlighted in bold, while second-best are underlined.}
\label{result_abla}
\resizebox{\linewidth}{!}{
    \begin{tabular}{ccc|cc}
    \toprule
    ALLM & LLM & Whisper & MMAU & MMAR\\
    \midrule 
    \multicolumn{5}{l}{\textit{Our Framework (w/ Proactive Iterative Document Refinement Loop)}} \\
    MiDashengLM-7B~\cite{dinkel2025midashenglm} & GPT-3.5-turbo~\cite{openai_gpt35turbo_finetune_2023} & Turbo & 67.30  & 49.72  \\
    MiDashengLM-7B~\cite{dinkel2025midashenglm} & GPT-4o~\cite{openai_gpt4o_system_card_2024} & Turbo & \underline{72.60}  & \textbf{58.85} \\
    Audio Flamingo 3~\cite{goel2025audio} & GPT-4o~\cite{openai_gpt4o_system_card_2024} & - & 69.40 & 55.36 \\
    Audio Flamingo 3~\cite{goel2025audio} & GPT-4o~\cite{openai_gpt4o_system_card_2024} & Turbo & \textbf{74.10}  & 55.80  \\
    Audio Flamingo 3~\cite{goel2025audio} & GPT-4o~\cite{openai_gpt4o_system_card_2024} & Large & 71.80  & \underline{57.24} \\
    Qwen2.5-Omni-3B~\cite{xu2025qwen2_5omni} & GPT-4o~\cite{openai_gpt4o_system_card_2024} & Turbo & 70.64  & 56.35 \\
    \midrule
    \multicolumn{5}{l}{\textit{Our Framework (w/o Proactive Iterative Document Refinement Loop)}}  \\
    MiDashengLM-7B~\cite{dinkel2025midashenglm} & GPT-4o~\cite{openai_gpt4o_system_card_2024} & / & 63.40 & 41.88\\
    Audio Flamingo 3~\cite{goel2025audio} & GPT-4o~\cite{openai_gpt4o_system_card_2024} & / & 68.90 & 44.09\\
    Qwen2.5-Omni-3B~\cite{xu2025qwen2_5omni} & GPT-4o~\cite{openai_gpt4o_system_card_2024} & / & 66.70 & 45.41 \\
    \bottomrule
    \end{tabular}
}
\vspace{-0.8em}
\end{table}

\subsection{Main Results}
\vspace{-0.5em}
\textbf{Comparison with SOTA Methods.}
Table \ref{result_mmau_mini} presents a comparison of AGR with SOTA audio reasoning methods on MMAU-mini. AGR not only surpasses open-source models but also outperforms the proprietary Gemini model, achieving the best performance. On MMAR (see Table \ref{result_mmar}), AGR significantly outperforms all open-source models and achieves results comparable to Gemini-2.0-Flash-Lite~\cite{vertexai_gemini_2_0_flash_lite_docs_2025}. Besides, our multi-agent framework yields substantial performance gains over direct inference with MiDahengLM, particularly on reasoning tasks involving speech and mixed audio types.

\textbf{Ablation Studies.}
The results of ablation studies are shown in Table \ref{result_abla}. A significant performance drop is observed when replacing GPT-4o~\cite{openai_gpt4o_system_card_2024} with GPT-3.5-turbo~\cite{openai_gpt35turbo_finetune_2023} in our iterative document refinement loop, particularly on the MMAR dataset. Since the LLM serves as the planning, interaction, and answering agent, its reasoning capability is a decisive factor in the final performance. We also replace our ALLM with Audio Flamingo 3~\cite{goel2025audio} (in configurations with and without the Whisper) and Qwen2.5-Omni-3B~\cite{xu2025qwen2_5omni}, which results in only slight performance variations. We infer the reason is that current ALLMs have comparable perceptual abilities, but their reasoning capabilities still differ significantly. Furthermore, the removal of our iterative document refinement loop causes a consistent performance drop for all tested ALLMs, most notably on the MMAR dataset. This confirms the effectiveness of our loop, which allows the model to continuously reflect on existing information, complete any missing evidence, and build a comprehensive evidence chain to support the final reasoning result.

\begin{table}[t]
\centering
\vspace{-0.8em}
\caption{Performance of different rounds on MMAU-mini and MMAR. Best results are highlighted in bold.}
\label{result_iteration}
\resizebox{\linewidth}{!}{
    \begin{tabular}{c|ccccc}
    \toprule
    Dataset & Iteration 0 & Iteration 1 & Iteration 2 & Iteration 3 & Iteration 4 \\
    \midrule 
    MMAU-mini & 68.90 & 72.90 & \textbf{73.80} & 71.80 & 71.90 \\
    MMAR & 44.09 & 54.59 & 56.35 & \textbf{57.24} & 57.02 \\ 
    \bottomrule
    \end{tabular}
}
\vspace{-0.9em}
\end{table}

\textbf{Effects of Iterative Rounds.}
We analyze the impact of the number of iterative rounds on model performance in Table \ref{result_iteration}. The initial iteration yields the most significant performance gain on both datasets, as it recovers the most critical missing information, validating our framework's effectiveness. Performance peaks at two rounds on MMAU-mini and at three rounds on MMAR, consistent with MMAR’s higher complexity and need for deeper exploration. With four rounds, performance drops on both datasets, likely because extra rounds introduce noise and irrelevant cues.




\section{Conclusion and Discussion}

In this work, we proposed AGR, a unified, training-free MAS that transforms audio deep reasoning into a text-based task. By decoupling perception from reasoning and employing a proactive iterative refinement loop, our framework synergizes the perceptual strengths of ALLMs with the advanced reasoning capabilities of LLMs. Experiments validate the effectiveness of this ``diagnose-plan-act" strategy, showing significant performance gains, particularly on high-level semantic tasks like speaker and content analysis.
Future work will focus on enhancing signal-level reasoning by developing more specialized evidence generators for low-level acoustic cues.

\section{Acknowledgments}
This work was supported by the National Natural Science Foundation of China (No. 62471420), GuangDong Basic and Applied Basic Research Foundation (2025A1515012296), and CCF-Tencent Rhino-Bird Open Research Fund.





\bibliographystyle{IEEEbib}
\bibliography{refs}

\begin{thebibliography}{10}

\bibitem{su2025audio}
Yi~Su, Jisheng Bai, Qisheng Xu, Kele Xu, and Yong Dou,
\newblock ``Audio-language models for audio-centric tasks: A survey,''
\newblock {\em arXiv preprint arXiv:2501.15177}, 2025.

\bibitem{ma2025mmar}
Ziyang Ma, Yinghao Ma, Yanqiao Zhu, Chen Yang, Yi-Wen Chao, Ruiyang Xu, Wenxi Chen, Yuanzhe Chen, Zhuo Chen, Jian Cong, et~al.,
\newblock ``Mmar: A challenging benchmark for deep reasoning in speech, audio, music, and their mix,''
\newblock {\em arXiv preprint arXiv:2505.13032}, 2025.

\bibitem{sakshimmau}
S~Sakshi, Utkarsh Tyagi, Sonal Kumar, Ashish Seth, Ramaneswaran Selvakumar, Oriol Nieto, Ramani Duraiswami, Sreyan Ghosh, and Dinesh Manocha,
\newblock ``Mmau: A massive multi-task audio understanding and reasoning benchmark,''
\newblock in {\em The Thirteenth International Conference on Learning Representations}, 2025.

\bibitem{liu2024caven}
Xiulong Liu, Sudipta Paul, Moitreya Chatterjee, and Anoop Cherian,
\newblock ``Caven: An embodied conversational agent for efficient audio-visual navigation in noisy environments,''
\newblock in {\em Proceedings of the AAAI conference on artificial intelligence}, 2024, vol.~38, pp. 3765--3773.

\bibitem{nie2024reason2drive}
Ming Nie, Renyuan Peng, Chunwei Wang, Xinyue Cai, Jianhua Han, Hang Xu, and Li~Zhang,
\newblock ``Reason2drive: Towards interpretable and chain-based reasoning for autonomous driving,''
\newblock in {\em European Conference on Computer Vision}, 2024, pp. 292--308.

\bibitem{xie2025audio}
Zhifei Xie, Mingbao Lin, Zihang Liu, Pengcheng Wu, Shuicheng Yan, and Chunyan Miao,
\newblock ``Audio-reasoner: Improving reasoning capability in large audio language models,''
\newblock {\em arXiv preprint arXiv:2503.02318}, 2025.

\bibitem{goel2025audio}
Arushi Goel, Sreyan Ghosh, Jaehyeon Kim, Sonal Kumar, Zhifeng Kong, Sang-gil Lee, Chao-Han~Huck Yang, Ramani Duraiswami, Dinesh Manocha, Rafael Valle, et~al.,
\newblock ``Audio flamingo 3: Advancing audio intelligence with fully open large audio language models,''
\newblock {\em arXiv preprint arXiv:2507.08128}, 2025.

\bibitem{ding2025kimi}
Ding Ding, Zeqian Ju, Yichong Leng, Songxiang Liu, Tong Liu, Zeyu Shang, Kai Shen, Wei Song, Xu~Tan, Heyi Tang, et~al.,
\newblock ``Kimi-audio technical report,''
\newblock {\em arXiv preprint arXiv:2504.18425}, 2025.

\bibitem{ghosh2024gama}
Sreyan Ghosh, Sonal Kumar, Ashish Seth, Chandra Kiran~Reddy Evuru, Utkarsh Tyagi, S~Sakshi, Oriol Nieto, Ramani Duraiswami, and Dinesh Manocha,
\newblock ``Gama: A large audio-language model with advanced audio understanding and complex reasoning abilities,''
\newblock in {\em Proceedings of the Conference on Empirical Methods in Natural Language Processing}, 2024, pp. 6288--6313.

\bibitem{elizalde2023clap}
Benjamin Elizalde, Soham Deshmukh, Mahmoud Al~Ismail, and Huaming Wang,
\newblock ``Clap learning audio concepts from natural language supervision,''
\newblock in {\em IEEE International Conference on Acoustics, Speech and Signal Processing}, 2023, pp. 1--5.

\bibitem{lipping2022clotho}
Samuel Lipping, Parthasaarathy Sudarsanam, Konstantinos Drossos, and Tuomas Virtanen,
\newblock ``Clotho-aqa: A crowdsourced dataset for audio question answering,''
\newblock in {\em IEEE European Signal Processing Conference}, 2022, pp. 1140--1144.

\bibitem{vertexai_gemini_2_5_flash_docs}
{Google Cloud},
\newblock ``Gemini 2.5 flash | generative ai on vertex ai,'' 2025.

\bibitem{vertexai_gemini20flash_docs_2025}
{Google Cloud},
\newblock ``Gemini 2.0 flash | generative ai on vertex ai,'' 2025.

\bibitem{vertexai_gemini_2_0_flash_lite_docs_2025}
{Google Cloud},
\newblock ``Gemini 2.0 flash-lite | generative ai on vertex ai,'' 2025.

\bibitem{dinkel2025midashenglm}
Heinrich Dinkel, Gang Li, Jizhong Liu, Jian Luan, Yadong Niu, Xingwei Sun, Tianzi Wang, Qiyang Xiao, Junbo Zhang, and Jiahao Zhou,
\newblock ``Midashenglm: Efficient audio understanding with general audio captions,''
\newblock {\em arXiv preprint arXiv:2508.03983}, 2025.

\bibitem{xu2025qwen2_5omni}
Jin Xu, Zhifang Guo, Jinzheng He, Hangrui Hu, Ting He, Shuai Bai, Keqin Chen, Jialin Wang, Yang Fan, Kai Dang, et~al.,
\newblock ``Qwen2. 5-omni technical report,''
\newblock {\em arXiv preprint arXiv:2503.20215}, 2025.

\bibitem{ghosh2025audio}
Sreyan Ghosh, Zhifeng Kong, Sonal Kumar, S~Sakshi, Jaehyeon Kim, Wei Ping, Rafael Valle, Dinesh Manocha, and Bryan Catanzaro,
\newblock ``Audio flamingo 2: An audio-language model with long-audio understanding and expert reasoning abilities,''
\newblock in {\em Forty-second International Conference on Machine Learning}, 2025.

\bibitem{openai_gpt4o_system_card_2024}
{OpenAI},
\newblock ``Gpt-4o system card,'' https://openai.com/index/gpt-4o-system-card, 2024,
\newblock Model snapshot used: gpt-4o-2024-08-06.

\bibitem{radford2023robust}
Alec Radford, Jong~Wook Kim, Tao Xu, Greg Brockman, Christine McLeavey, and Ilya Sutskever,
\newblock ``Robust speech recognition via large-scale weak supervision,''
\newblock in {\em International conference on machine learning}, 2023, pp. 28492--28518.

\bibitem{openai_gpt35turbo_finetune_2023}
{OpenAI},
\newblock ``Gpt-3.5 turbo fine-tuning and api updates,'' https://openai.com/index/gpt-3-5-turbo-fine-tuning-and-api-updates, Aug. 2023.

\end{thebibliography}

\end{document}